\documentclass[a4paper,12pt]{article}
\usepackage{amsmath,amssymb}
 \usepackage{epsfig} \usepackage{floatflt}
\usepackage{wrapfig}  \usepackage{subfigure}
\newcommand{\beq}{\begin{equation}}
\newcommand{\eeq}{\end{equation}}    %
\def\eV{\relax\ifmmode{\rm e\kern-0.12em V}\else{\rm e\kern-0.12em V{ }}\fi}
\def\MeV{\relax\ifmmode{\rm M\eV}\else{\rm M\eV}\fi}
\def\GeV{\relax\ifmmode{\rm G}\eV\else{\rm G\eV}\fi}      
\def\gbar{\relax\ifmmode{\bar{g}}\else{$\bar{g}${}}\fi}
 \newcommand{\ns}{\normalsize} 
 \newcommand{\fns}{\footnotesize}
\def\gbar{\relax\ifmmode{\bar{g}}\else{$\bar{g}${}}\fi}

\begin{document}
\begin{flushright}
{\small\it Dedicated to the memory \\ of Albert 
 Tavkhelidze}\end{flushright} \smallskip
 \begin{center}
\centerline{\sf\large Coupling running through the
     Looking-Glass of dimensional Reduction} \bigskip

  D.V. Shirkov \medskip

 {\small Bogoliubov Lab., JINR, Dubna}
 \end{center}
\begin{abstract}
 The dimensional reduction, in a form of transition 
 from four to two dimensions, was used in the 90s in a 
 context of HE Regge scattering.Recently, it got a new 
 impetus in quantum gravity where it opens the way to 
 renormalizability and finite short-distance behavior.

 We consider a QFT model $g\,\varphi^4\,$ with running
 coupling defined in both the two domains of different
 dimensionality; the $\gbar(Q^2)\,$ evolutions being  
 duly correlated at the reduction scale $\,Q\sim M.$ 
 Beyond this scale, in the deep UV 2-dim region, the 
 running coupling does not increase any more. Instead, 
 it {\it slightly decreases} and tends to a finite value 
 $\gbar_2(\infty)\,<\,\gbar_2(M^2)\,$ from above. 
 
 As a result, the global evolution picture looks quite 
 peculiar and propose a base for the modified scenario 
 of gauge couplings behavior with UV fixed points 
 provided by dimensional reduction instead of 
 leptoquarks.\end{abstract}

 \section{\large Introduction}
 \subsection{\ns Motivation}
  Initial motive for considering the topic mentioned
 in the title is related to the hot quest of the Higgs
 particle still escaping from direct observation.
 Anticipating the possibility that no Higgs peak will be
 observed in the ``window'' $140\pm 25\,\GeV\,$, we try
  to investigate the {\it Ginzburg-Landau-Higgs} \
 possibility, with Higgs field being a classical field
 with nonzero constant component $\sim 250\,\GeV\,$
 sufficient for the mass production in the current
 version of SM.  In other words, we are inclined to use
 the fact that the Higgs mechanism for creating masses
 could work when the Higgs field is not a quantum one
 being rather an analog of the Ginzburg-Landau
 two-component order parameter from the theory of
 superconductivity\footnote{For details see our recent
 overview \cite{60let}}.%%

   However, changing a quantum Higgs for the classical
 external field yields the trouble of renormalization in
 the EW sector of SM. Having no intention to enter this
 complicated problem we prefer to postpone its solution
 and look for some temporary practical remedy for the
 time being\footnote{Here, on can rely upon the
 well-known examples :\\ -- In the early 30s, Dirac was
 brave enough\cite{Dirac33} to use cutoff on the proton
 mass in discussing the momentum dependence of electron
 charge $e(Q)\,,$ a prototype of the QED running
 coupling.\\
 -- In the 50s and 60s, after the devising of the
 renormalization procedure, phenomenologists still
 widely and fruitfully used hardly non-renormalizable
 4-fermion interaction {\it \`a l\'a Fermi} for
 analyzing weak interactions. They postponed the
 problem solution till reaching the so called
 ``unitary limit'' at $\,W_{c.m.}\sim$100\,\GeV.}.
 The possible way is to involve an invariant
 regularization procedure, like the Pauli-Villars one.
 To this goal, it is more intriguing to exploit a
 transition from the four-dimensional manifold to the
 one with a \ {\it smaller} \ number of dimensions
 $\,d<4\,$ at high enough energy or small distance.
  Technically, this could provide us with artificial 
 cutoff with only one additional parameter, the range 
 of reduction.\smallskip

 The trick with changing the number of dimensions is a
 frequent one in current literature (on superstrings
 ect.) devoted to the HE behavior. This transition a l\'a
 Kaluza-Klein to a larger number of dimensions $\,D> 4\,$
 inevitably confronts us with the non-renormalizability.
  Instead, we consider another, a rather opposite
 possibility, the {\it dimensional reduction} (DR). 

  To explore some practical aspects of the DR, as a 
 first test-flight to terra incognita, we turn here to
 a limited subject, the issue of transferring the
 renormalization-invariant running coupling
 $\,\gbar(Q^2)\,$ through the region of reduction and
 relating its behavior in two domains with different
 dimensionality. %%  
 
 \subsection{\ns Dimensional reduction.}
 The dimensional reduction was used first about 15 
 years ago\footnote{See paper \cite{irina} and 
 references therein.} as a pragmatic tool in the 
 analysis of the HE Regge scattering. This line of 
 reasoning was refreshed \cite{Dejan10} quite recently 
 in the context of the LHC physics with a more explicit 
 emphasis of the DR physical implementation. 
 In the last decade it became quite popular in quantum 
 gravity. Here, a class of models has been devised by 
 Horava (see, e.g., paper \cite{horava} and references 
 therein) with asymptotic anisotropy between space and 
 time dimensions in the short distance UV limit 
 \footnote{This activity was motivated by remarkable 
 observation by Ambjorn, Loll and others\cite{ambjorn} 
 in the causal dynamical triangulation approach to quantum 
 gravity on lattice. See also an overview \cite{nieder07}.}. 

 Our attitude do not imply any modification of the 
 special relativistic concept of the time. We just have 
 in mind some smooth reduction of the spatial topological 
 dimensions. Probably this scenario is akin   one 
 another approach formulated recently\cite{fractal10} 
 for the quantum field living in the fractal spacetime.
 \smallskip

 {\small\bf Agreement on DR.} One can
 discuss the mechanism of dimensional reduction either
 in the space-time terms or in the energy-momentum ones
 using the presumptive assumption that
 \begin{quote}{\it Reduction at the space-time scale \
 $x_{dr}\sim 1/M_{dr}\,$ is, in a sense, equivalent to
 the reduction at the energy-momentum scale} \
 $\,p_{dr}\sim M_{dr}\,.$ \end{quote}

 This tentative agreement will be used below to compare
 the Lagrangian (space-time) approach with the direct 
 {\it ad hoc} \ modification of the momentum integration
 of Feynman integrals.
  In the course of the first approach we use a sharp
 conjunction as an approximation to a softer mechanism
 of a continuous DR in the second one. \smallskip  

 {\ns\bf Classical illustration.} To illustrate the idea
 of approximation, imagine a wine bottle (e.g., posed
 vertically) like one presented on Fig.1(a). It 
 consists of the main cylindrical body $B$ with a 
 relatively large radius $R$ and a length $L\,.$ The 
 bottle's neck $N$ of length $l\,$ and smaller radius 
 $\,r$ is connected with the main part by a ``collar''
  -- a narrowing transition region $\,C\,$ of a
 varying radius and a short length $\,l_{coll}\,.$
 \begin{figure}[h]
 \centerline{\includegraphics[width=0.70\textwidth]
 {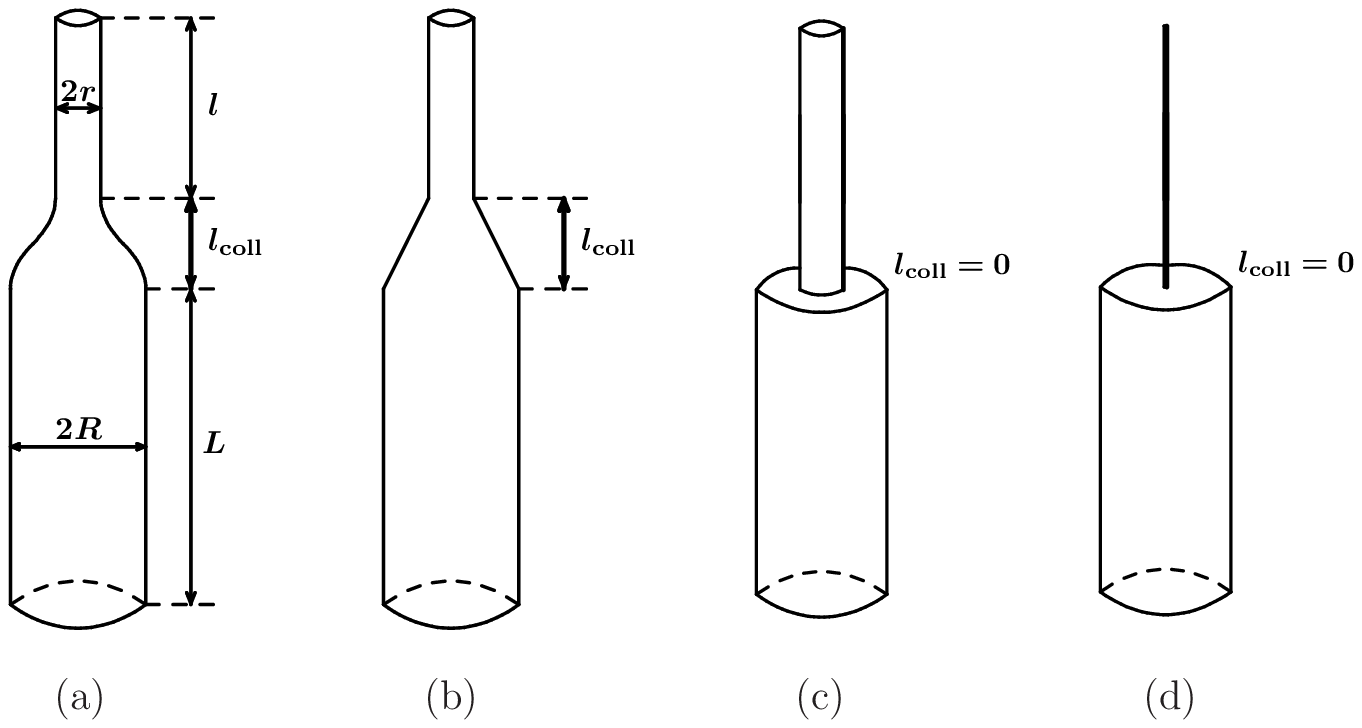}}
 \vspace*{1mm}
 {\fns \caption \ \ 
 (a)- Wine-bottle with a long body and a long neck;\
  (b) - Bottle with a conical collar: \\
 (c) - Bottle with a horizontal collar: \
 (d) - Bottle with a horizontal collar and a 
 thread-like  neck.}
 \end{figure}

 Suppose now that there is a mathematical model for 
 some process on the two-dimensional surface $S_2\,$ 
 of the bottle. One can have in mind a stationary 
 boundary value problem for the second order partial 
 differential equation(s), or some dynamical process 
 like wave propagation with radiation or heat 
 conductivity and so on.

 A number of problems with exact analytical solutions 
 on the surfaces $S_{R,L}\,$ and $S_{r,l}\,$ of 
 cylindrical parts can be found. Many of them could 
 be solved for the whole two-dimensional manifold
 $\,S_2=S_{R,L}+S_{r,l}+S_{coll}\,$ for simple enough
 smooth forms of the junction collar region $C\,.$

  Of particular interest are non-stationary processes,
 like a solitary wave propagating upward due to some
 short-time perturbation at the lower edge of the bottom.

  It will be instructive to study several issues : \\
 --  the dependence of solution details on the form of
  the surface $S_{coll}$ of the collar region; \\
 -- transition to a sharp change of radius (to a
    horizontal collar) as \ $l_{coll}\to 0$, Fig. 1(c);\\
 -- the limiting case $\,r\to 0\,,$ that is a transition
  from the 2-dim surface of the neck $S_{r,l}$ to the
 1-dim linear manifold $S_{0,l} \to L_l\,,$  Fig. 1(d).

 In the further analysis, along with smooth transition 
 in Section 2.1, we shall use ``hard conjunction''in 
 Section 2.2 of two regions with different dimensions. 
 The first one resembles the Fig.1(a) in the limit 
 $r\to 0\,,$ while the second one is an analog of a 
 system presented on Fig.1(d).

 \section{\large Effective \gbar \ for the 
  $\,\varphi^4\,$ model in various dimensions}     %%s2

  Take the one-component scalar massive quantum field
 $\varphi(x)\,$ with the self-interaction Lagrangian
 \beq \label{eq1}
  L=T-V;\quad V(m,g;\varphi)=\frac{m^2}{2}\,\varphi^2
 + \frac{4\pi^{d/2} M^{4-d}}{9}\,g_d\,
 \varphi^4\,; \quad\quad  g > 0\,\eeq
 in parallel in four ($d=4\,$) and two ($d=2\,$) 
 dimensions.

 Limit ourselves to the one-loop approximation level for
 \gbar \ that corresponds to the only Feynman diagram
 contribution, the first correction to the 4-vertex
 function, Fig.2.

 \begin{center}
  \begin{figure}[!ht]\label{fig3}   \centering
 \epsfig{scale=0.60,figure=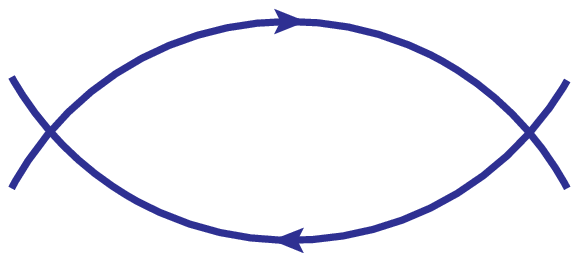} \vspace{-4mm}
 \caption{\fns  One-loop vertex diagram}
 \end{figure} \end{center}

 Its contribution \ $I$ \ enters into the effective
 running coupling as follows:
  \begin{equation}\label{eq2} \gbar(q^2)=
\frac{g_{i}}{1-g_{i}\,I\,(q^2; m^2, m_i^2)}\,.\eeq

 \subsection{\ns Smooth \ DR \ in the momentum picture}
  We start with reduction of dimensions in Feynman 
 integral by modifying metrics in the 
 momentum space \vspace{-4mm}
\beq\label{d-4}
 d k=d^4\,k\to d_Mk\,=\frac{d^4\,k}{1+k^2/M^2}\,;
 \quad k^2=\mathbf{k}^2 - k_0^2\,.\eeq
 In particular, for the one-loop integral Fig.2 \ 
 one gets  {\small  \[ 
 I\left(\frac{q^2}{m^2}\right)=\frac{i}{\pi^2}
 \int\frac{dk}{(m^2+k^2)[m^2+(k+q)^2]}\,\to\,\frac{i}
 {\pi^2}\,\int\frac{d_Mk}{(m^2+k^2) [m^2+(k+q)^2]}
  =J(\kappa;\mu)\,,\] } %%  \eeq
 with \ $\kappa=q^2/4\,m^2,\,\mu=M^2/m^2, \ q^2=
 \mathbf{q}^2-q_0^2\,.$ The integral $J$ can be 
 calculated explicitly. We give its asymptotics. In the 
 \ ``deep 4-dim'' region $m^2\ll q^2\ll M^2$ one gets 
 an ``intermediate'' \ logarithmic behavior with the 
 $M$ playing the role of the Pauli-Villars regulator. 
 Meanwhile, in the ``deep 2-dim'' \ region 
 $q^2\gg M^2\gg m^2\,,$ the UV limit is finite.
 In usual normalization  \vspace{-4mm}

 \[ J \to J_i =J(q^2/4m^2;\mu)- J(m_i^2/4\,m^2;\mu)\,;
 \quad m_i \sim m\,, \] 
 one has \vspace{-4mm}

 \begin{equation}\label{uv-2} J^{[4]}_i(\kappa;\mu)\,
 \sim \ln\left(\frac{q^2}{m_i^2}\right)\,;\quad
 J^{[2]}_i(\kappa;\mu)\,\sim\ln\left(\frac{4\,M^2}
 {m_i^2}\right)+\frac{M^2}{q^2}\ln\frac{q^2}{M^2}\,.\eeq

  The first expression is rising; while the second,
 decreasing. The maximum value of $J$ is attained at the 
 DR scale $q^2\sim M^2\,$ and is close to
 $\ln(M^2/m^2)\,.$ Hence, due to the DR, the 
 $\gbar(q^2)\,$  evolution changes drastically. The 
 effective coupling slightly diminishes\footnote{This 
 reverse evolution is not seen in the common massless, 
 pure logarithmic, RG analysis.} beyond the reduction 
 scale and tends to a finite value, see Fig.3.
  \begin{figure}[!ht] \centering
 \epsfig{scale=0.85,figure=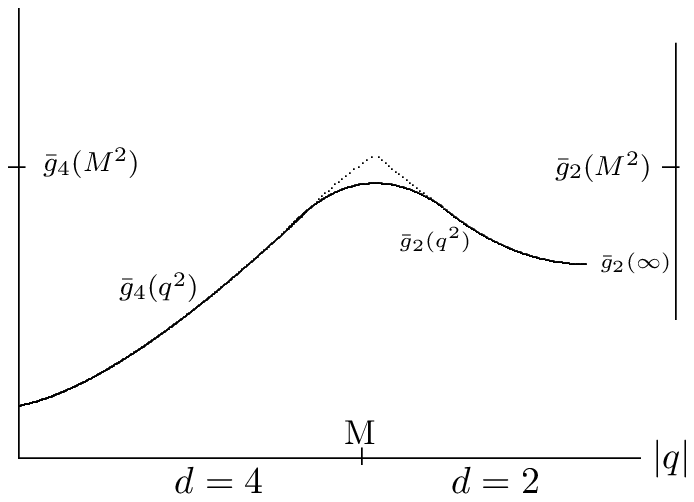}
 \caption{\fns The  effective coupling evolution for
 the $\varphi^4\,$ model with DR.} \end{figure}

 \subsection{\ns Reduction with Lagrangians}
 Now, along with the {\it Agreement on DR}, return to
 the Lagrangian description eq.(\ref{eq1}) in terms
 of fields. There, under transition to the $d=2\,$ case,
 the field $\varphi_4(x)$ loses its dimensionality 
 $\varphi_4(x)\to\varphi_2(x)\sim M^{-1}\,\varphi_4(x)
 \,,$ while the coupling constant (as in eq.(1)) acquires 
 it: $g_4\to\,M^2\,g_2,$ with some parameter $\,M\,$ that 
 we put equal to the DR scale $\,M=M_{dr}\,.$ Below, the 
 dimensionless constant $g_2\,$ will be used.
 
 \subsubsection{\ns\it Invariant coupling in $d=2\,$.} 
 \ In two dimensions, one can as well use finite Dyson
 transformations, formulate RG invariance and 
 mass-dependent renormalization group\footnote{As it 
 was introduced in the mid-50s in \cite{nc56}: see also 
 Chapter VIII in monograph \cite{book} or Chapter IX,
 \S\,51 in its 3rd edition\cite{book2} and Appendix 9 
 in the textbook \cite{QF}.}, define a ``massive'' 
 running coupling with its explicit one-loop solution
 \cite{blank56}(see, also \S 43.1 in \cite{book} 
 \footnote{Unhappily, this piece was omitted in the 
 next edition \cite{book2}.} and paper \cite{mrg92})
 {\small \beq\label{eq5} 
 \gbar^{[2]}(q^2)=\tfrac{g}{1-g\,M^2\, I_2(q^2/m^2)}
 =\tfrac{g}{1-g\, J_2(q^2/m^2)}\,;\,\, J_2\left(
 \frac{q^2}{m^2}\right)=\frac{i}{\pi}\int\frac{M^2\,\,
 d^2k}{(m^2+k^2)\,[m^2+(k+q)^2]}\,.\eeq}
 Here, $\,J_2\,$ is a finite one-loop contribution from
 the 4-vertex diagram, Fig.2, in two dimensions. It is a
 positive monotonously decreasing function.   
 Asymptotically,  $J_2\sim(M^2)/q^2\,\ln(q^2/m^2)\,,$ 
  -- just like in the second eq.(\ref{uv-2}). 
 Therefore, two-dimensional effective coupling in the
 UV limit tends to its limiting fixed value from above.

 \subsubsection{\ns\it Hard conjunction at the
  reduction scale}                             % 2.2.3
 To obtain the joint picture of coupling evolution, one
 has to consider a transition from the ``low-energy''
 4-dim region \ $q^2 < M^2\,$ to the ``high-energy'' 
 2-dim one \ $q^2 > M^2\,.$

 For the ``hard'' conjunction\footnote{In the classical
 wine-bottle model, this hard conjunction corresponds
 to Fig.1(d).}, the continuity property 
 $\gbar_4\,(M^2)=\gbar_2(M^2)=g_M\,$ yields
 \begin{equation}\label{uv-4}
 \gbar_4(q^2)=\frac{g_M}{1-g_M\,\ln(q^2/M^2)}\,;
 \quad  q^2 \leq M^2 \, \end{equation}
 and, along with expression (\ref{eq5}),
 \begin{equation}\label{gbar2}
 \gbar_2(q^2)=\frac{g_M}{1-g_M\,\left[J_2(q^2/m^2)- 
 J_2(M^2/m^2)\right]}\,; \quad q^2 \geq M^2 \, \eeq
 with finite UV limit
 \begin{equation}\label{infty2}              %%  eq 8
 \gbar_2(\infty)=\frac{g_M}{1+g_M\, 
 J_2(M^2/m^2)}\,\, <\,g_M\,.\eeq
  This means that above the reduction scale the 
 effective coupling evolves down to its final UV 
 limit. This behavior corresponds to Fig.3.

 \section{\large Discussion}  %% sec3

 In the above analysis, one more alternative to the
 standard Higgs mechanism within the Standard Model was
 considered. The main idea consists in employing the
 possibility of reducing the number of dimensions in the
 far UV limit, the possibility that is intensively
 discussed now in the context of quantum gravity. \\
 Taking for definiteness the reduction from common four
 dimensions to two dimensions at some high enough scale
 $|q|\sim M\,,$ we studied the issue of effective
 coupling behavior for the $\varphi^4\,$ scalar model
 (\ref{eq1}). Using hard conjunction, by continuity: \
 $\gbar_4(M^2)=\gbar_2(M^2)\,$  just at the DR scale,
 we got the joint picture which is very close to the \
 $\gbar\,$ behavior in the case of continuous DR by 
 changing metric in the momentum space via eq.(\ref{d-4}). 
 The resulting picture is presented on Fig.3. \medskip

  Its essential technical feature is the change of the 
 familiar logarithmic growth of the \ $\gbar_4\,$ 
 running coupling for the slight decrease of the 
 $\,\gbar_2(Q^2)\,$ beyond the reduction scale. There, 
 in the effective 2-dim region, the \ $\varphi^4_2\,$ 
 interaction is super-renormalizable and the {\it \, 
 one-loop mass-dependent contribution to running
 coupling is a decreasing function.} \, Due to this, 
 the $\gbar(Q^2)$ evolution changes its pattern. At 
 the infinity it tends to its limiting value from 
 above, as in eq.(\ref{gbar2}).

  This peculiar property, the reverse running coupling
 evolution to a fixed point beyond the \ ``distorted
 mirror'' marked by the reduction scale, has a chance 
 to upgrade the Grand Unification scenario with the 
 $M\,$ value being of the order or even greater than 
 the hypothetical lepto-quark scale --
   see qualitative illustration on the Fig.4.
 
 Indeed, a rough estimate reveals that difference 
 between $1/\gbar(\infty)$ and $1/g_M$ is of order of 
 unity; being imported to the GUT context, it could 
 produce effect numerically sufficient to close the 
 famous discrepancy triangle\cite{GUT}. 

  \begin{figure}[!ht] \centering
 \epsfig{scale=0.65,figure=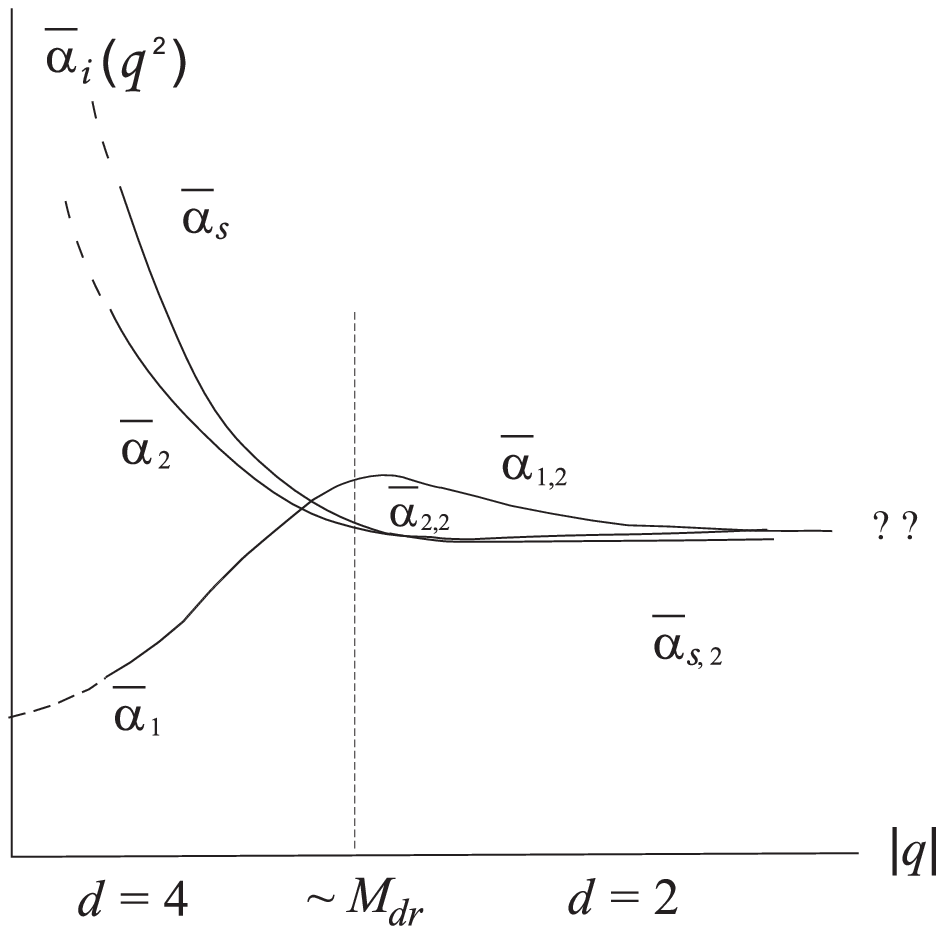} 
 \caption{\fns Modified scheme of the hypothetical 
 Great Unification provided by dimensional
  reduction  instead of leptoquarks.}
 \end{figure} 
 
  The notable observation is that the change of 
 geometry could yield the same final result as an 
 explicit change of dynamics (by adding leptoquark 
 fields etc.).

  Among further quests that are in order, let us put 
 in the first place the issue of examining the chance 
 of detecting some physical signal \ ``through the
 looking-glass at scale $M$''\ that would provide us 
 with direct evidence on the existence of dimension 
 reduction of any kind. It could be fruitful to study 
 some classical problems formulated at the end of 
 Section 1 to enrich our intuition. \bigskip 

   \centerline{\large\sf  Acknowledgments}\smallskip

  It is a pleasure to thank Drs. Irina Aref'eva and 
 Plamen Fiziev for useful discussions. The courtesy 
 of Dr. Jean Zinn-Justin for informing on relevant 
 literature is also appreciated. This research has 
 been partially supported by the presidential grant 
 Scientific School--3810.2010.2 and
 by RFFI grant 08-01-00686. \bigskip
 
 \centerline{\sf\large Appendix}
 \small Few relevant integrals. \medskip
 
 {\small
 1. \ First, an auxiliary one-loop Feynman integral 
 (of the Fig.2 type) with different masses (see 
 eq.(DVI.31) on page 341 of the textbook~\cite{QF} 
 American edition)
 \begin{equation} f(q^2)=
 f(\tfrac{q^2}{M^2};\mu^2) \sim \frac{i}{\pi^2}
 \int\frac{d^4k}{(m^2+k^2)(M^2+(k+q)^2)}\,;\quad 
 q^2=\mathbf{q^2}-q_0^2\,.  \end{equation}
 According to eq.(DVI.32), at $\mu^2= m^2/M^2\ll 1$ 
 the result of integration subtracted at zero, 
 $f_0(0)=0\,$   looks like  
 \[f_0\left(\frac{q^2}{M^2};\mu^2\right)=\frac{K(q)}
 {2}\ln\frac{M^2+q^2+q^2\,K}{M^2+q^2-q^2\,K}+\frac{1}
 {2}\left(1+\frac{M^2}{q^2}\right)\ln\frac{M^2}{m^2}-
 1+ o\left(\frac{m^2}{M^2}\right);\]
  with \vspace{-4mm}
 \[K(q)=\left[\left(1+\tfrac{(M+m)^2}{q^2}\right)
 \left(1+\tfrac{(M-m)^2}{q^2}\right)\right]^{1/2}\sim 
 1+\tfrac{M^2}{q^2}-\tfrac{2m^2M^2}{q^2(q^2+M^2)}+
 o\left(\tfrac{m^2}{M^2}\right)\,.\]
 
   Start first with intermediate logarithmic 
 asymptotics at $ m^2\ll q^2\ll M^2\,.$ There,
 \begin{equation}
 f_0^{UV}\left(\frac{q^2}{M^2}\right)\simeq\left(1+
 \frac{M^2}{q^2}\right)\,\ln\frac{M^2+q^2}{M^2}-1+
 o\left(\frac{m^2}{q^2}\right).\end{equation} 
 This asymptotic expression still satisfies the 
 normalization condition $\,f_0^{UV}(0)=0\,.$}\medskip 
 
 2. \ Another 4-dim integral in the dimensionless 
 normalization {\small
 \begin{equation}
  J(\kappa;\mu)=\frac{i}{\pi^2}\,\int\frac{M^2
 \,\,dk}{(m^2+k^2)[m^2+(k+q)^2](M^2+k^2)}\,\eeq
 can be presented via $f_0$ in the simple form
 \beq J(\kappa;\mu)=\frac{M^2}{M^2-m^2}\left[f_o
 \left(\tfrac{q^2}{m^2};1\right)-f_o\left(
 \tfrac{q^2}{M^2};\mu^2\right)\right]\,.\eeq 
   
  It is evident that at the ``D4 asymptotic region''
 $m^2\ll q^2\ll M^2\,$ the first term $f_0(q^2/m^2)$ 
 with its rising logarithmic asymptotics dominates, 
 while beyond the looking glass in the D2 UV region
 $\,q^2\gg M^2\,$ the whole expression tends to 
 finite limit.} \medskip
  
  3. For the very end, one more expression resulting 
 from the 2-dim integral (5) in the Euclidean 
 domain at \ $q^2 >0$ \ and \ $\,k^2=k_1^2-k_0^2\to
 k_E^2=k_1^2+k_4^2 > M^2$
 \[J_2\left(\frac{q^2}{m^2}\right)=\frac{i}{\pi}
 \int_M\,\frac{M^2\,d^2k}{(m^2+k^2)\,[m^2+(k+q)^2]}
 =\frac{1}{\pi}\int_M\,\frac{M^2\,d^2_E\,k}
 {(m^2+k_E^2)\,[m^2+(k+q)_E^2]}\,; \] 
 \[J_2(0)= \frac{M^2}{m^2+M^2}\,;\quad 
 J_2\left(\frac{M^2}{m^2}\right)=\]  \medskip
 
 4. Numerical estimate. Due to eqs.(8),(13), the
  difference 
 \[\frac{1}{\gbar(\infty)}-\frac{1}{g_M}=
 J_2(M^2/m^2)\]
 is of order of unity. Accordingly, being imported 
 to the GUT context, it can produce effect numerically 
 sufficient to close the famous discrepancy triangle.

 \end{document}